\documentclass[aps,prl,twocolumn,superscriptaddress,nofootinbib]{revtex4-2}
\usepackage[english]{babel}
\usepackage{bm}
\usepackage{amsmath}
\usepackage{amssymb}
\usepackage{textcomp}
\usepackage{dutchcal}
\usepackage[dvips]{graphicx}
\usepackage[colorlinks=true, allcolors=cyan]{hyperref}
\usepackage{diagbox}
\usepackage{multirow}
\usepackage{threeparttable}
\usepackage{xfrac}
\usepackage{xcolor}

\begin{document}
\title{Geometric oscillations of local Hall and Nernst effects in ballistic graphene at weak magnetic fields}
\author{Z.~Z.~Alisultanov}
\email{zaur0102@gmail.com}
\affiliation{Abrikosov Center for Theoretical Physics, MIPT, Dolgoprudnyi, Moscow Region 141701, Russia}
\affiliation{Institute of Physics of DFRS of the RAS, Makhachkala, 367015, Russia}
\author{A.~V.~Kavokin}
\email{kavokinalexey@gmail.com}
\affiliation{Abrikosov Center for Theoretical Physics, MIPT, Dolgoprudnyi, Moscow Region 141701, Russia}
\affiliation{School of Science, Westlake University, 18 Shilongshan Road, Hangzhou 310024, Zhejiang Province, China}

\begin{abstract}
We predict a novel class of magnetotransport oscillations in ballistic graphene specific for a ring-shape geometry. Using the Büttiker-Landauer formalism, we analytically obtain the local Hall and Nernst coefficients in the weak-field ballistic regime. These coefficients exhibit pronounced oscillations as functions of both the magnetic field and the angular positions of the measurement probes. The oscillations originate from the discrete set of skipping orbits that geometrically connect the contacts, with resonances occurring when the angular separation between contacts times the radius of the disk equals an integer number of cyclotron diameters. Unlike conventional quantum oscillations in conductivity, this effect is robust at room temperature and can dominate local thermoelectric signals. This geometric control of ballistic flow provides a platform for studying electron hydrodynamics and engineering phase-coherent devices, with potential applications in sensitive terahertz detectors and thermal management systems.
\end{abstract}

\maketitle


The exploration of graphene's exceptional electronic properties has unveiled a plethora of novel transport phenomena, with ballistic conduction standing as a hallmark of its high quality and weak electron-phonon coupling~\cite{Novoselov2005, Geim2007}. In the ballistic regime, charge carriers traverse micron-scale distances without scattering, enabling the study of fundamental quantum effects and paving the way for ultra-high-speed electronic devices~\cite{Du2008, Katsnelson2006a}. This coherent transport is primarily governed by the material's unique linear Dirac-like band structure, which leads to unusual phenomena such as Klein tunneling and Veselago lensing~\cite{Katsnelson2006a, Cheianov2007}. The electronic response of ballistic graphene junctions has been extensively characterized through quantum conductance measurements, revealing a wealth of information about its intrinsic properties and potential for post-silicon electronics~\cite{Russo2008, Miao2007}.

Ballistic transport in a magnetic field is a powerful probe of Fermi surface geometry and spin-orbit coupling. Transverse magnetic focusing (TMF), where carriers on skipping orbits are focused onto a collector, reveals spin splittings and topological features~\cite{van1989coherent,beconcini2016scaling,rendell2023spin,taychatanapat2013electrically,berdyugin2020minibands}. Thus, ballistic graphene serves not only as a high-mobility platform but also as a sensitive diagnostic tool for complex spin-orbit phenomena.

\begin{figure}[h!]
    \centering
    \includegraphics[width=0.9\linewidth]{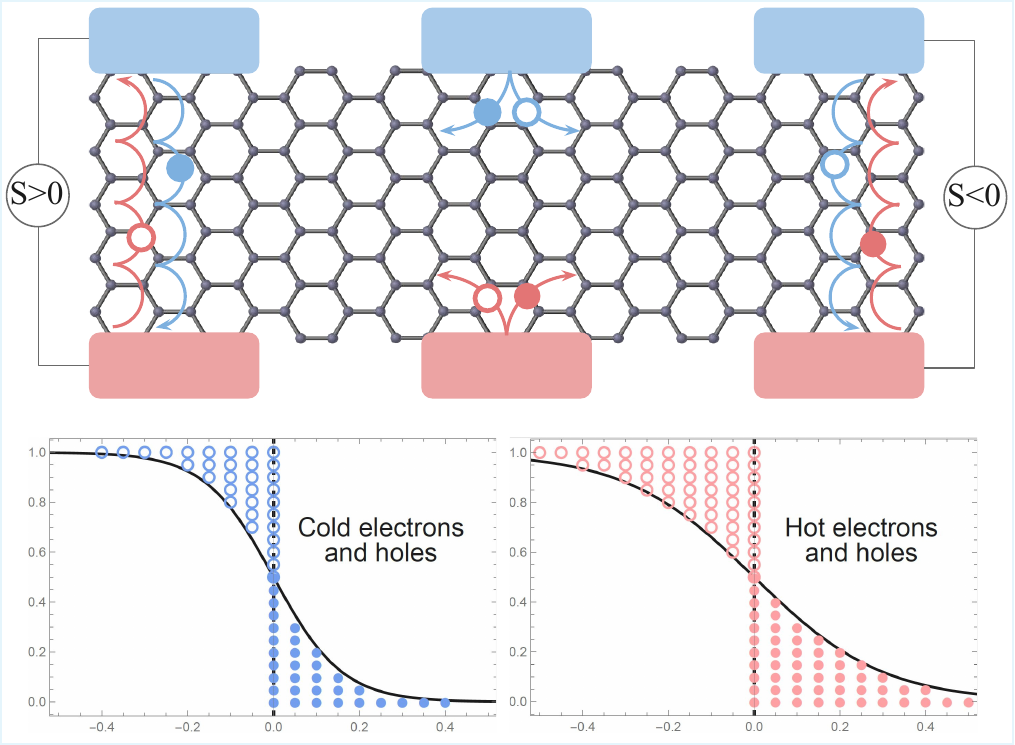}
    \caption{(a) Ballistic graphene sheet connected to top (cold, blue) and bottom (hot, red) reservoirs. Electrons (filled circles) and holes (empty circles), driven by the temperature gradient, are deflected oppositely by perpendicular magnetic field B, producing spatially separated edge currents. (b) Fermi-Dirac distributions for undoped graphene at low and high temperatures. The hot reservoir generates a larger carrier population than the cold one. Magnetic field separates these carriers spatially, yielding a pronounced Nernst effect and a finite local Seebeck voltage at each edge. Crucially, the local Seebeck coefficient has opposite signs on opposite edges.}
    \label{sketch}
\end{figure}

The sensitivity of ballistic trajectories to the Fermi surface geometry, exploited so successfully in TMF, finds a natural counterpart in the thermoelectric domain. Thermoelectric measurements have emerged as a unique probe of correlated electronic states in two-dimensional systems. Thermopower studies in monolayer graphene mapped fractional quantum Hall states, extracting energy gaps and identifying even-denominator states via entropic sensitivity~\cite{sultana2026thermopower}. This demonstrates thermoelectric transport as a sensitive spectroscopic tool. Beyond fundamental probes, the terahertz-driven Nernst effect in graphene moiré superlattices enables high-sensitivity, bias-free detection for next-generation optoelectronics~\cite{elesin2025enhanced}.

\begin{figure}
    \centering
    a)\includegraphics[width=0.4\linewidth]{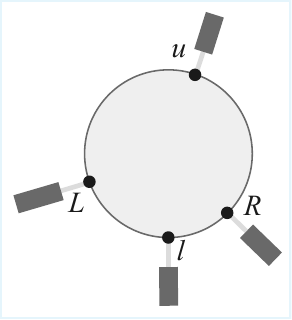}~
    b)\includegraphics[width=0.45\linewidth]{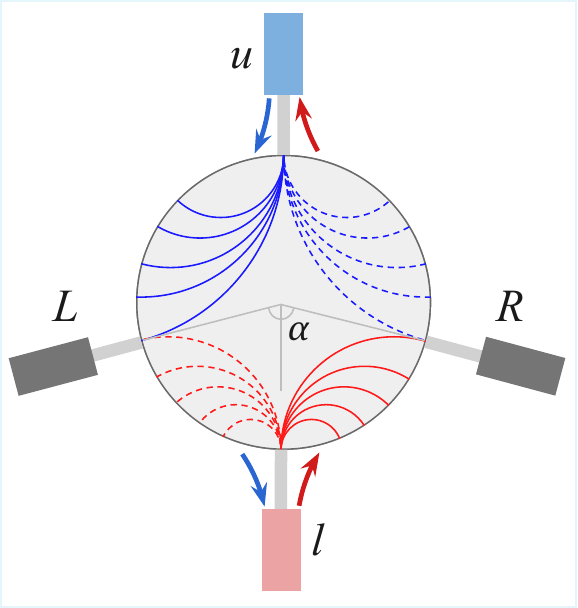}

    ~

    ~
    
    ~~\includegraphics[width=0.4\linewidth]{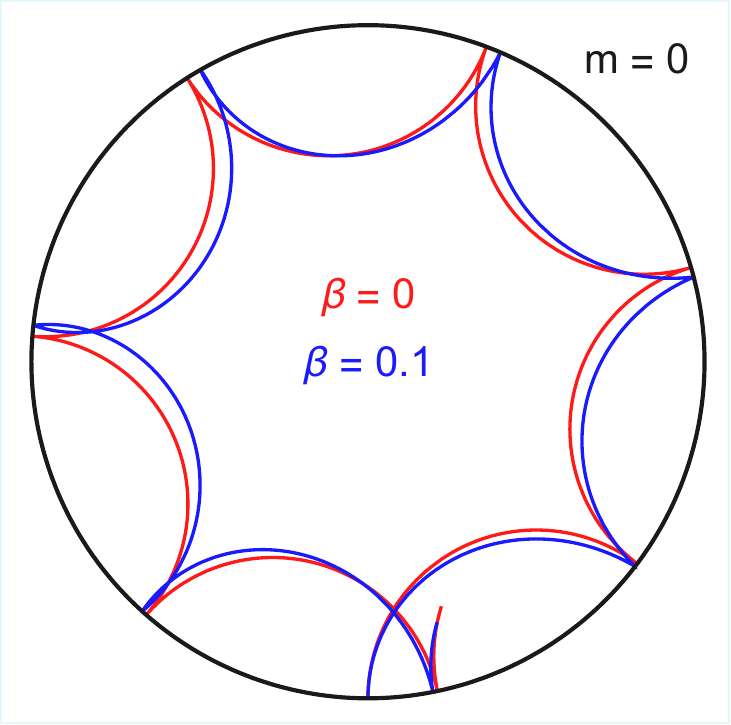}
    \includegraphics[width=0.4\linewidth]{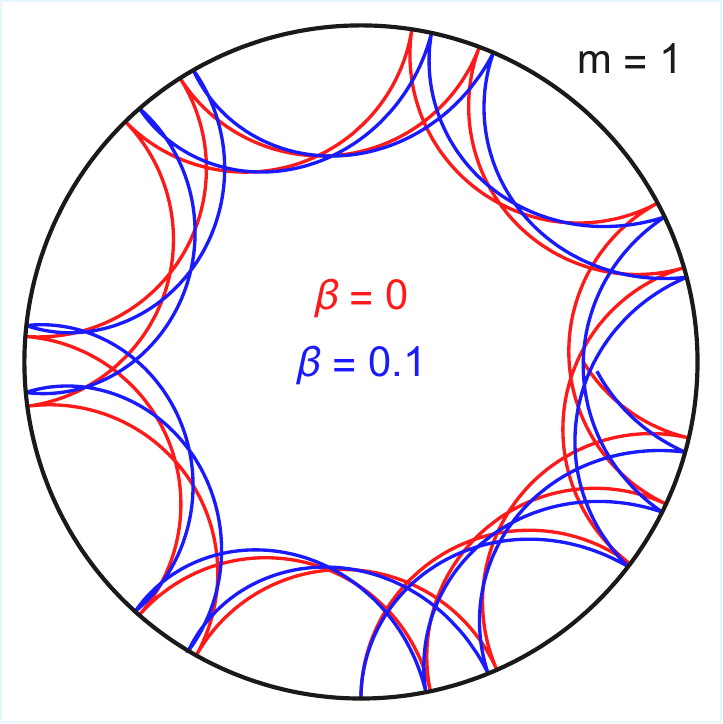}
    
    c)\includegraphics[width=0.4\linewidth]{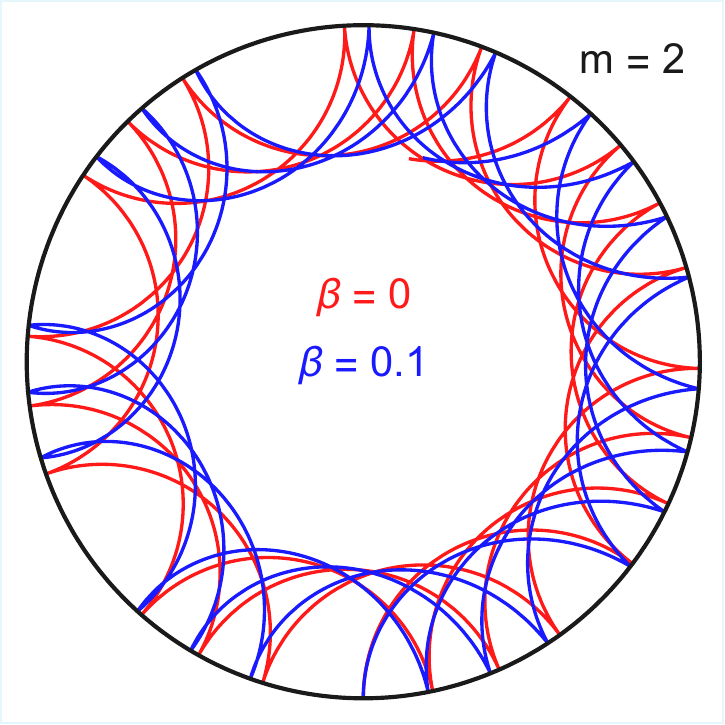}
    \includegraphics[width=0.4\linewidth]{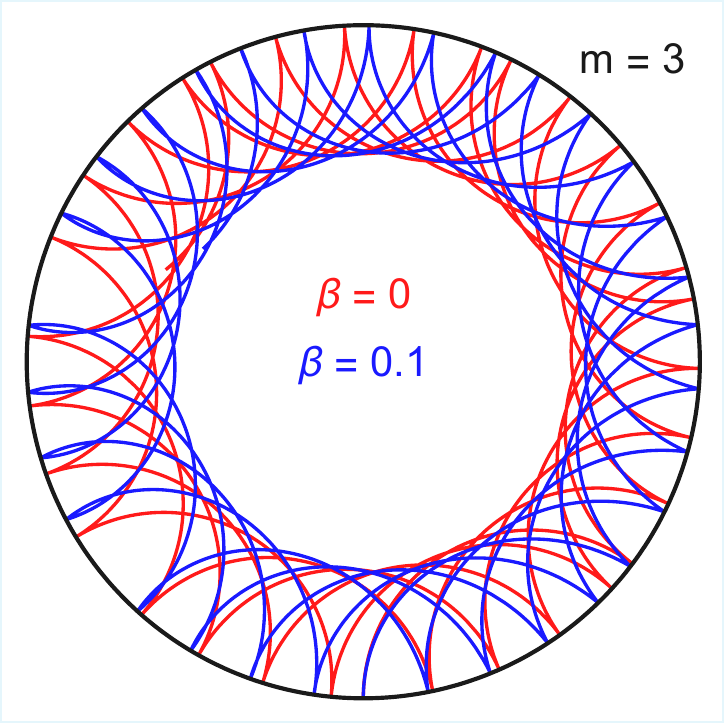}
    \caption{(a) Four-terminal Büttiker setup for ballistic disk transport measurements. Contacts l, R, u, L are arbitrarily positioned at the edge, enabling local electrical (longitudinal, Hall) and thermoelectric (Seebeck, Nernst) measurements via applied bias imbalances. (b) Configuration used in our calculations. (c) Simulated electron trajectories in a graphene disk for $\beta = E/(\upsilon_F B)=0$ (red) and $0.1$ (blue). Panels show trajectories completing 1–4 full revolutions along the edge. $\beta = 0.1$ corresponds to a strong electric field regime experimentally accessible with typical laboratory fields ($\beta \sim 0.01-0.1$).}
    \label{general sketch}
\end{figure}

In this Letter, we study the local classical Hall and Nernst effects in a ballistic graphene disk (see Fig.~\ref{general sketch}) in a weak magnetic field using the Büttiker–Landauer formalism. The disk geometry enables an exact analytical calculation of local transport characteristics. We find that the local Hall and Nernst coefficients oscillate both with magnetic field and with the position of the measuring contacts. These oscillations are purely geometric in origin and emerge in weak magnetic fields, that is in the classical regime.

To begin with, and to qualitatively illustrate the nontrivial character of local electronic transport in the ballistic regime, we consider a system whose schematic is shown in Fig.~\ref{sketch}. Since the transport of carriers is ballistic, electrons travel without bulk scattering — scattering occurs only at the boundaries — and the current is determined solely by the imbalance between electrons emitted from and absorbed into the reservoirs. We herein address the intermediate regime characterized by the cyclotron radius $r_c$ being smaller than the sample length $L$ ($r_c < L$), where skipping orbits are present yet quantum effects remain negligible. We demonstrate that the local Seebeck coeffitient exhibits opposite signs at opposite sample edges. A qualitative illustration of this effect is provided in Fig.~\ref{sketch}.

Our preliminary analysis focuses on charge-neutral (undoped) graphene, where the total Seebeck coefficient vanishes. At finite temperature, charge carriers—both electrons and holes—are activated near the Fermi level. The respective carrier densities are given by
\begin{gather}
n_e\left(T\right) = \int_{0}^{\infty} f(\varepsilon, T) \rho(\varepsilon) d\varepsilon, \\
n_h\left(T\right) = \int_{-\infty}^{0} \left[1 - f(\varepsilon, T)\right] \rho(\varepsilon) d\varepsilon,
\end{gather}
where $f(\varepsilon, T)$ is the Fermi-Dirac distribution function and $\rho(\varepsilon)$ is the density of states of reservoirs.

In the presence of a perpendicular magnetic field, the Lorentz force deflects electrons and holes in opposite directions. Consequently, their skipping orbits propagate along opposite edges of the sample. This results in two counter-propagating edge channels: an electron channel on one edge and a hole channel on the other, carrying current in the same direction relative to the sample (see Fig.~\ref{sketch}). Therefore, at each end of the sample, two distinct reservoirs inject carriers into these edge channels: one reservoir sources electrons and the other sources holes.

We now calculate the ballistic current between reservoirs held at different temperatures. Considering the right edge in Fig.~\ref{sketch}, the bottom reservoir injects an electron stream, while the top reservoir injects a hole stream. The net current at the right edge is thus given by
\begin{gather}
I_{R} = - e \bar{\upsilon}_{\mathrm{dr}} (n_e^{T'} + n_h^T) = - e \bar{\upsilon}_{\mathrm{dr}}\times\nonumber\\ \left[\int_{0}^{\infty} f_{T'}(\varepsilon) \rho(\varepsilon) d\varepsilon
+ \int_{-\infty}^{0} \left[1 - f_T(\varepsilon + eV)\right] \rho(\varepsilon) d\varepsilon\right],\label{current right}
\end{gather}
where $T'=T+\Delta T$,  $\bar{\upsilon}_{\mathrm{dr}}$ is the average drift velocity of the carriers (the explicit form for this quantity does not play a role for qualitative analysis), $e$ is the elementary charge, and $V$ is the voltage developing between the contacts. Similarly, we can write an expression for $I_L$. Although local currents are non-zero, the net current through the entire system must, of course, vanish. This implies that the net thermopower in undoped graphene is zero, a result that has been repeatedly confirmed experimentally by the vanishing of the Seebeck coefficient at the charge neutrality point. 

In the linear approximation from Eq.~\eqref{current right} (and a similar formula for $I_L$) we can find the corrections to the currents $\Delta I_{L,R}$, due to the temperature difference. In the open-circuit regime, the net currents vanish: $\Delta I_{R}=\Delta I_{L}=0$. Solving these equations for the induced voltage yields the local Seebeck coefficients
\begin{gather}
S_R = -S_L = V/\Delta T.
\end{gather}

This simple example demonstrates that local thermoelectric transport in a ballistic system can exhibit an intriguing feature: the thermopower can have opposite signs at different ends of the system. This sign reversal of the local Seebeck coefficient at opposite edges is a direct manifestation of counterpropagating ballistic currents carried by electrons and holes, in agreement with the recent analysis of nonlocal thermoelectric response in graphene~\cite{kavokin2024anomalous}. Next, we will examine electronic and thermoelectric transport in ballistic graphene within the more general framework of the Büttiker–Landauer theory. In particular, we will show that the sign reversal effect occurs also for the local Nernst effect that exhibits non-trivial geometric oscillations.

We begin by constructing a general formalism for calculating electronic and thermoelectronic transport in a two-dimensional ballistic system. To this end, we employ a four-terminal setup (see Fig.~\ref{general sketch}a) and the Landauer-Büttiker formalism~\cite{landauer1957spatial,buttiker1986four,meir1992landauer,note1}. For the sake of analytical treatment, we consider a system of a disk-like geometry. We will show below that in such a geometry, the calculation of ballistic trajectories reduces to a purely geometric problem. Each terminal consists of a reservoir (the wide, dark regions), characterized by its respective temperature and chemical potential, and a quasi-one-dimensional ballistic lead (the narrow, light regions). The terminals are labeled as l (lower), R (right), u (upper), and L (left).

Following Büttiker~\cite{buttiker1986four}, we write the current injected from the \(i\)-th reservoir (\(i=l,R,u,L\)) into the system (considered positive) as
\begin{gather}
I_i=\frac{e}{h}\int_{-\infty}^{\infty}d\varepsilon\sum_{j\neq i}\mathcal{T}_{ji}\left(\varepsilon\right)\left(f_i\left(\varepsilon\right)-f_j\left(\varepsilon\right)\right),\label{Buttiker equation}
\end{gather}
where $\mathcal{T}_{ji}$ denotes the transmission probability from contact $i$ to contact $j$. We now proceed using this formula. First, we note that \( I_i = 0 \) in the equilibrium case where \( f_i = f_j = f_0 \). Furthermore, in the ballistic limit, we can compute the coefficients \( \mathcal{T}_{ij} \) explicitly. Evidently, these coefficients can only take values of 0 or 1, as no stochastic scattering processes are present.

We employ Eq.~\eqref{Buttiker equation} to analyze electrical and thermoelectrical transport. For instance, we can study the Hall current \(I_{LR}^{H}\) between terminals \(L\) and \(R\) by setting \(\mu_l=\mu_0+eV, \mu_u=\mu_0, \mu_L=\mu_R\) and \(T_l=T_u=T_R=T_L\). Within the linear response approximation, we obtain
\begin{gather}
I_{LR}^{H}=\frac{e^2}{h}\frac{\tilde{\mathcal{T}}_{uL}\left(\tilde{\mathcal{T}}_{lL}+\tilde{\mathcal{T}}_{lR}\right)-\tilde{\mathcal{T}}_{lL}\left(\tilde{\mathcal{T}}_{uL}+\tilde{\mathcal{T}}_{uR}\right)}{\tilde{\mathcal{T}}_{uL}+\tilde{\mathcal{T}}_{uR}+\tilde{\mathcal{T}}_{lL}+\tilde{\mathcal{T}}_{lR}}V,
\label{Hall current}
\end{gather}
where $\tilde{\mathcal{T}}_{ij}=\int_{-\infty}^{\infty}\mathcal{T}_{ij}\left(\varepsilon\right)\left(-\frac{\partial f}{\partial \varepsilon}\right)d\varepsilon$. 

Similarly, we can study the Nernst current \(I_{LR}^{N}\) between terminals \(L\) and \(R\) by setting \(T_u=T_R=T_L=T,~T_l=T+\Delta T\) and \(\mu_l=\mu_u+\left(\partial\mu/\partial T\right) \Delta T\)
\begin{gather}
I_{LR}^N=\frac{e}{h}\frac{\Lambda_1 \partial\mu/\partial T+\Lambda_2}{\tilde{\mathcal{T}}_{uL}+\tilde{\mathcal{T}}_{uR}+\tilde{\mathcal{T}}_{lL}+\tilde{\mathcal{T}}_{lR}}\Delta T,
\label{Nernst current}
\end{gather}
where $\Lambda_1=\tilde{\mathcal{T}}_{uL}\tilde{\mathcal{T}}_{lR}-\tilde{\mathcal{T}}_{uR}\tilde{\mathcal{T}}_{lL}$,
$\Lambda_2=\left(\tilde{\mathcal{T}}_{uL}+\tilde{\mathcal{T}}_{lL}\right)\tilde{\tilde{\mathcal{T}}}_{lR}-\left(\tilde{\mathcal{T}}_{uR}+\tilde{\mathcal{T}}_{lR}\right)\tilde{\tilde{\mathcal{T}}}_{lL}$,
$\tilde{\tilde{\mathcal{T}}}_{ij}=\int_{-\infty}^{\infty}\mathcal{T}_{ij}\left(\varepsilon\right)\frac{\varepsilon-\mu}{T}\left(-\frac{\partial f}{\partial \varepsilon}\right)d\varepsilon$.
The Eqs.~\eqref{Hall current},\eqref{Nernst current} are key for our next calculations. 

To compute the transmission coefficients \( \mathcal{T}_{ij} \) explicitly, we begin with the model illustrated in Fig.~\ref{general sketch}b. A point reservoir injects charge carriers (electrons and holes) into the system, which then propagate along skipping orbits of different cyclotron radii under the applied magnetic field. Our objective is to calculate all carrier trajectories that connect the reservoirs to the measurement contacts. As an illustrative example, consider carriers emitted from the lower reservoir \(l\). In the presence of a perpendicular magnetic field, electrons injected from this reservoir will be deflected toward the right contact \(R\), while holes will be deflected toward the left contact \(L\) (see Fig.~\ref{general sketch}b). Consequently, counter-propagating charge and heat currents flow from reservoir \(l\) toward the two adjacent contacts. A similar analysis applies to carriers emitted from any other reservoir. The self-consistent treatment of all these processes, combined with the open-circuit condition for each contact, yields the full set of kinetic coefficients and allows us to compute the local Hall and Nernst responses.

The cyclotron radius in graphene is given by
\begin{gather}
r_c=\frac{\varepsilon}{\upsilon_F eB},
\end{gather}
We will take advantage of the following useful property of intersecting circles: the angles of intersection between two circles are equal at both points of their intersection. In the context of our problem, this means that the particle's launch angle from the reservoir uniquely determines all subsequent angles of incidence and reflection at the boundary. This is the fundamental geometric reason for the analytical predictability of the skipping orbits that connect the reservoirs to the measurement contacts.

For simplicity, we first assume that charge carriers are emitted from the reservoirs at an angle of $\pi/2$ relative to the disk edge (later, we generalize this to the case of arbitrary emission angles). The problem then reduces to calculating all currents flowing into and out of each contact (reservoirs $l,u$ and measurement probes $L,R$). The positions of all contacts on the disk adge are defined by their respective angular coordinates. For instance, with respect to the bottom reservoir (red in Fig.~\ref{general sketch}b), under symmetric arrangement, the measurement probes are located at an angle $\alpha$ for the right probe and an angle $2\pi - \alpha$ for the left one. For the top reservoir (blue in Fig.~\ref{general sketch}b), the corresponding angle is $\pi$.

Consider an electron emitted normally from the bottom reservoir. If the cyclotron radius is $r_c$ and the disk radius is $R$, the angular distance $\phi$ to the point of first collision with the edge is given by
\begin{gather}
\phi=2\arctan\frac{\varepsilon}{\upsilon_FeBR}.\label{angle distance}
\end{gather}
Clearly, the condition for such an electron to reach the right measurement probe $R$ can be generally expressed as
\begin{gather}
\alpha+2\pi m=n\phi,\label{angle condition}
\end{gather}
where $n=1,2,\ldots$ is a positive integer equal to the number of skipping events along the edge before the electron reaches the contact. Trajectories with $m=0$ connect the bottom reservoir directly to the right contact. Trajectories with $m=1$ start at the bottom reservoir, complete one full revolution along a disk edge, and terminate at the right contact, and so forth. Next, we can neglect the trajectories with $m \neq 0$. Clearly, such trajectories are characterized by velocities smaller than those for $m = 0$ and yield negligible contributions to the current. The magnitude of these contributions decreases with the increase of $m$. Next, we will assume $m=0$. Indeed, numerical estimates of the contributions from trajectories with \(m > 0\) to the quantities \(\tilde{\mathcal{T}}_{ij}\), \(\tilde{\tilde{\mathcal{T}}}_{ij}\) show that already for \(m = 1\) these contributions are negligibly small: \(\tilde{\mathcal{T}}_{ij}, \tilde{\tilde{\mathcal{T}}}_{ij} \ll 1\) for \(m > 0\).

For an arbitrary electron emission angle $\theta$, the angle $\phi$ is given by
\begin{gather}
\phi(\theta) = 2\arctan\!\left(\frac{r_c \sin\theta}{R - r_c \cos\theta}\right).
\label{eq:phi_theta}
\end{gather}
At $\theta = \pi/2$, Eq.~\eqref{eq:phi_theta} reduces to Eq.~\eqref{angle distance}. When calculating kinetic coefficients, it is necessary to integrate over $\theta$ in the range $[0, \pi]$, taking into account the weighting factor $\sin^2\theta$ that accounts for the angular distribution.

\begin{figure}
    \centering
    \includegraphics[width=0.85\linewidth]{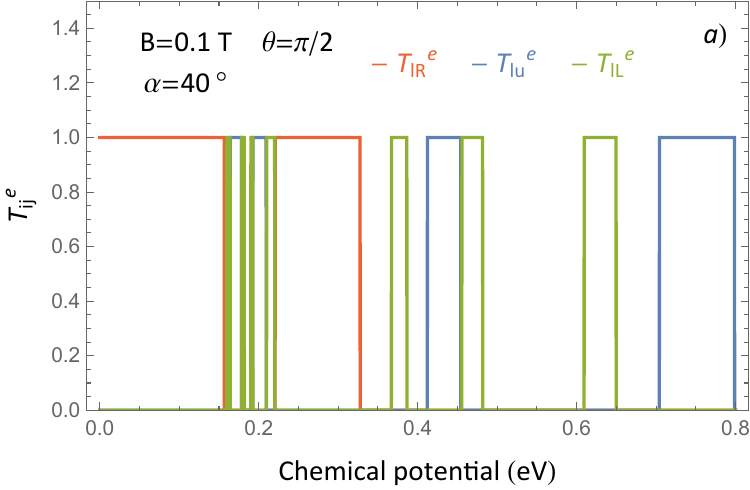}
    \includegraphics[width=0.85\linewidth]{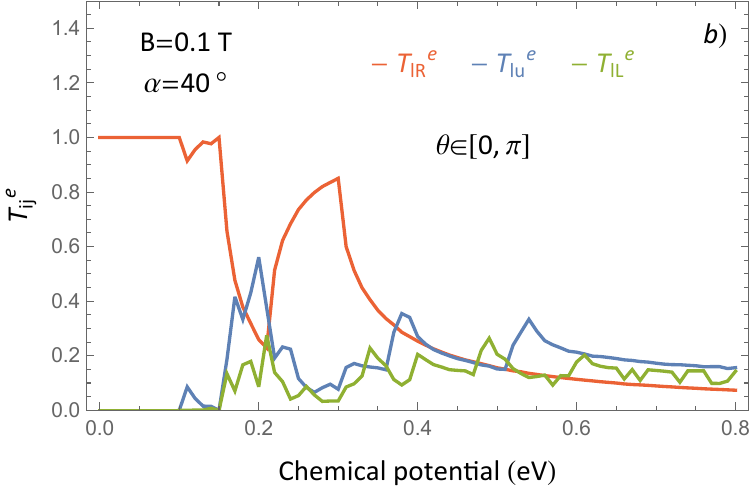}
    \includegraphics[width=0.85\linewidth]{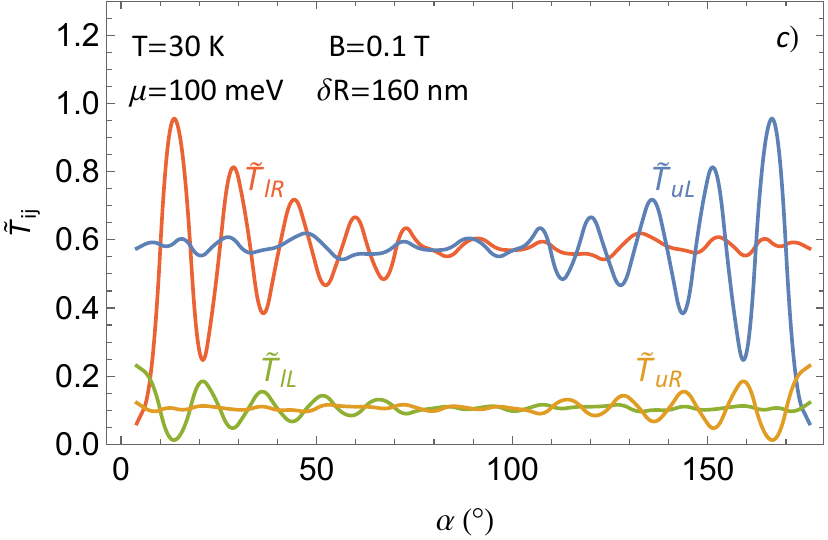}
    \caption{Dependencies of the values \(\mathcal{T}_{lR}^e\), \(\mathcal{T}_{lu}^e\), and \(\mathcal{T}_{lL}^e\) on the chemical potential at \(T = 0\) for a fixed emission angle \(\theta = \pi/2\) (a) and for all emission angles \(\theta \in [0,\pi]\) (b). We put $R=2.5~\mu\text{m}$ and $\delta R=315~\text{nm}$, where $\delta R=R\delta\alpha$ is the width of contacts. The panel (c) shows the angular dependences of the quantities $\tilde{\mathcal{T}}_{lR},~\tilde{\mathcal{T}}_{lL},~\tilde{\mathcal{T}}_{uR},~\tilde{\mathcal{T}}_{uL}$ at $\mu=100~\text{meV}$ and $\delta R=160~\text{nm}$. 
}
    \label{trans coeff}
\end{figure}

\begin{figure*}
    \centering
    \includegraphics[width=0.45\linewidth]{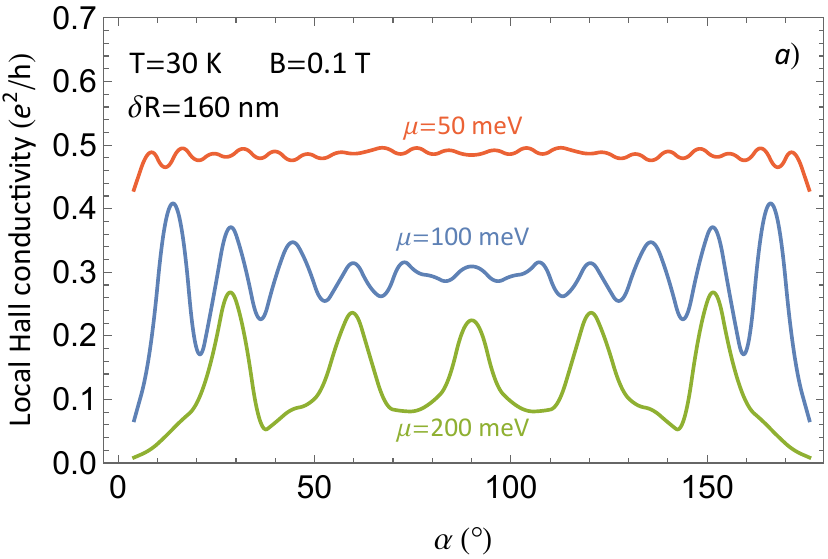}~~
    \includegraphics[width=0.45\linewidth]{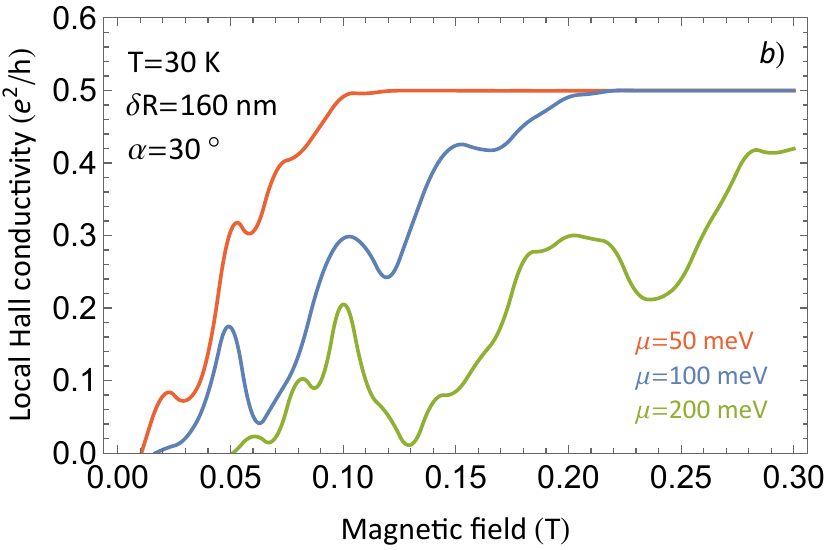}
    \caption{Local Hall conductivity as a function of the contact position at fixed magnetic field (a) and vs. field at fixed contact position (b). For $\mu < 50~\text{meV}$, conductivity is $e^2/2h$—a plateau with virtually no oscillations. As field increases, oscillations give way to plateaus.}
    \label{Hall oscillations}
\end{figure*}
 
Given condition~\eqref{angle condition}, we write the expression for \( \mathcal{T}_{ij} \) in the following form (into account the finite thickness of the contacts).
\begin{gather}
\mathcal{T}_{ij}\left(\varepsilon\right)=\frac{2}{\pi}\prod_{k=i+1}^{j-1}\sum_{n=1}^{\infty}\int_0^\pi \sin^2\theta d\theta\left(1-\mathcal{T}_{ik}\left(\varepsilon\right)\right)\nonumber\\
\times\Theta\left(\alpha_{ij}+\delta\alpha-n\phi\right)\Theta\left(n\phi-\alpha_{ij}+\delta\alpha\right).
\end{gather}
where $\alpha_{lR}=\alpha,~\alpha_{lu}=\pi,~\alpha_{lL}=2\pi-\alpha,~...$ and $\delta\alpha$ is a small angle corresponding to the angular width of the contacts, $\mathcal{T}_{ij}=\mathcal{T}_{ij}^e+\mathcal{T}_{ij}^h$. Then we obtain the following explicit expressions for $\mathcal{T}_{ij}^{e/h}$
\begin{gather}
\mathcal{T}^e_{lR}=\mathcal{T}^h_{lL}=\mathcal{T}_{\alpha}^B\left(\varepsilon\right),\\
\mathcal{T}^e_{lu}=\mathcal{T}^h_{lu}=\left(1-\mathcal{T}_{\alpha}^B\left(\varepsilon\right)\right)\mathcal{T}_{\pi}^B\left(\varepsilon\right),\\
\mathcal{T}^e_{lL}=\mathcal{T}^h_{lR}=\left(1-\mathcal{T}_{\alpha}^B\left(\varepsilon\right)\right)\left(1-\mathcal{T}_{\pi}^B\left(\varepsilon\right)\right)\mathcal{T}_{2\pi-\alpha}^B\left(\varepsilon\right),
\end{gather}
with analogous expressions for other contacts and $\mathcal{T}_{\alpha}^B\left(|\varepsilon|\right)=\frac{2}{\pi}\int_0^\pi \sin^2\theta d\theta
\sum_{n=1}^{\infty}\Theta\left(\alpha+\delta\alpha-n\phi\right)\Theta\left(n\phi-\alpha+\delta\alpha\right)$.

Fig.~\ref{trans coeff}(a,b) show the energy dependence of the transmission coefficients \(\mathcal{T}_{lR}^{e}\), \(\mathcal{T}_{lu}^{e}\), and \(\mathcal{T}_{lL}^{e}\) at \(\alpha=40^{\circ}\) and zero temperature, calculated for a fixed emission angle \(\theta = \pi/2\) (a) and for all emission angles \(\theta \in [0,\pi]\) (b). One can see that, at zero temperature these quantities take values of either 0 or 1.
At non-zero temperature, the sharp jumps are smoothed out, and the very narrow regions are strongly suppressed.

Fig.~\ref{trans coeff}c shows the angular dependences of the quantities $\tilde{\mathcal{T}}_{lR},~\tilde{\mathcal{T}}_{lL},~\tilde{\mathcal{T}}_{uR},~\tilde{\mathcal{T}}_{uL}$ at the chemical potential $\mu=100~\text{meV}$ and contact width $\delta R=160~\text{nm}$. First, it should be noted that at $\mu=0$ all quantities are identical and equal to 1/2. This follows from electron-hole symmetry. Next, we observe that for $\mu<50~\text{meV}$, except for the special case of the undoped system $\mu=0$, $\tilde{\mathcal{T}}_{lL},~\tilde{\mathcal{T}}_{uR}$ are almost zero. This means that the so‑called nearest‑neighbor contact approximation works well, where the main contribution to transport comes from trajectories leading from a given reservoir to the nearest neighboring one (right or left depending on the doping type: electron or hole). In the next paragraph we discuss this approximation and its experimental feasibility in detail. Finally, as the doping increases ($>50~\text{meV}$) or the contact width decreases, oscillations emerge that are of purely geometric origin. Peaks in these oscillations occur when the condition $\alpha=\phi\left(\mu\right)$ is satisfied (see Eq.~\eqref{angle condition}).

\begin{figure*}
    \centering
    \includegraphics[width=0.45\linewidth]{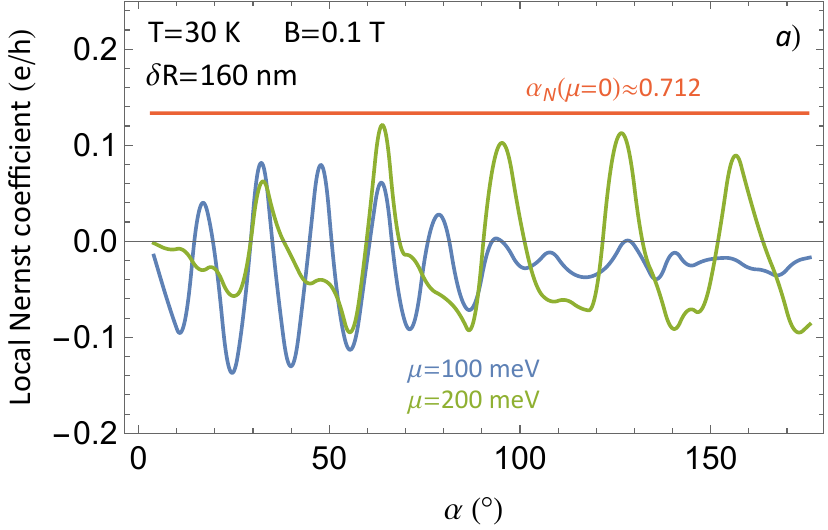}~~
    \includegraphics[width=0.45\linewidth]{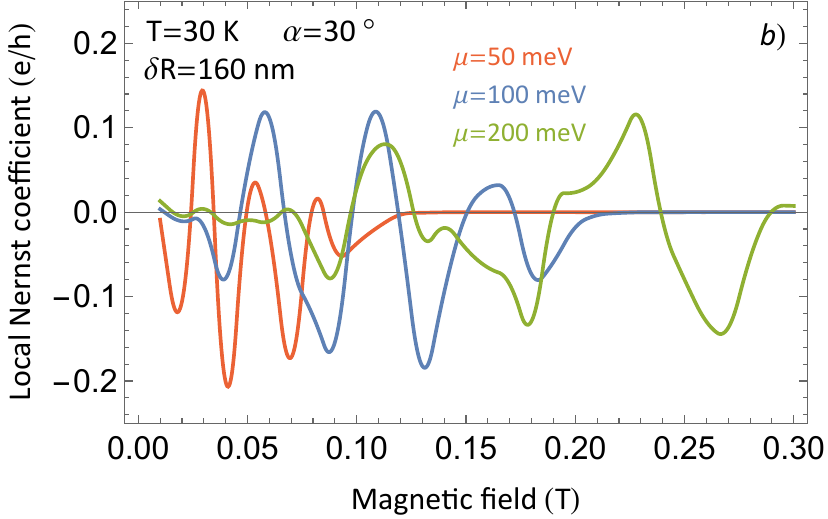}
    \caption{Local Nernst coefficient as a function of the contact position at fixed field (a) and vs. field at fixed contact position (b). At the charge neutrality point, the Nernst effect is maximal, with $\alpha_N(\mu=0)\approx0.712$. With doping, oscillations emerge around zero.}
    \label{Nernst oscillations}
\end{figure*}

We can consider a special case that significantly simplifies the calculations (we already mentioned this approximation in the previous paragraph). Suppose the contacts have a finite width $\Delta l$. It is then straightforward to see that all trajectories with $0 < 2r_c < \Delta l$ will reach the nearest contact. Some trajectories with $2r_c > \Delta l$ will also reach it, but their number is significantly smaller. This implies that electrons with energies $0 < \varepsilon < \upsilon_F e B \Delta l / 2c $ behave in this manner. Furthermore, note that the main contribution to the currents comes from the energies within the interval $[\mu - T, \mu + T]$. Therefore, in the undoped case, the main contribution (for a given carrier type) comes from energies $0 < \varepsilon < T$. Contributions from other states are exponentially suppressed. Consequently, if $\Delta l > 2c T / (\upsilon_F e B)$, then all relevant trajectories reach the nearest contact. We can thus state that the coefficients $T_{ij}$ equal 1 for the nearest contacts and 0 for all others. Hence, for this special case we have:
\begin{gather}
\mathcal{T}_{lR}^e=\mathcal{T}_{Ru}^e=\mathcal{T}_{uL}^e=\mathcal{T}_{Ll}^e\approx1,\\
\mathcal{T}_{ij}^e=0,~\text{for other i,j},
\end{gather}
etc.  
In realistic graphene samples, the contact width may on the order of 0.5 to 1 $\mu\text{m}$. Meanwhile, the quantity $2c T / (\upsilon_F e B)$ at a temperature of $100~K$ and a magnetic field $B=1~T$ is $10$ nm (100 nm for $B=0.1~\text{T}$
). Once the condition $\Delta l \gg 2c T / (\upsilon_F e B)$ is satisfied, this approximation is fully justified. In the doped case, the validity condition for this approximation can be written as: $T \ll \mu < e \upsilon_F B \Delta l/2$. For example, at $\Delta l = 500~\text{nm}$, we obtain $\mu < 0.25 \cdot B~(\text{eV})$. At $n = 10^{11}~\text{cm}^{-2}$, we have $\mu \approx 0.037~\text{eV}$. Therefore, our approximation is valid for $B > 0.15~\text{T}$. At $n = 10^{12}~\text{cm}^{-2}$, our approximation is valid for $B > 0.5~\text{T}$.

The observable features of these geometric resonances are manifested in both the electrical and thermoelectric response. Fig.~\ref{Hall oscillations} displays the local Hall conductivity as a function of the contact position \( \alpha \) at a fixed magnetic field [panel (a)] and as a function of the magnetic field at a fixed contact position [panel (b)]. At low doping (\( \mu < 50 \) meV), the Hall conductivity exhibits a robust plateau at \( e^2/2h \), indicating that the nearest-neighbor contact approximation holds and the skipping orbits are effectively "washed out" by the thermal spread. However, as the doping increases (e.g., \( \mu = 100 \) meV), sharp geometric oscillations emerge. These oscillations are particularly pronounced in the field dependence [Fig.~\ref{Hall oscillations}b], where the condition \( \alpha = \phi(\mu) \) selects discrete cyclotron radii that perfectly bridge the source and detector. Notably, at higher fields these oscillations give way to plateaus, marking the transition to a regime where the cyclotron radius becomes smaller than the contact width, effectively restoring the nearest-neighbor transport.

Fig.~\ref{Nernst oscillations} extends this analysis to the thermoelectric domain, revealing the local Nernst coefficient dependence on the contact positions. At the charge neutrality point (\( \mu = 0 \)), the Nernst effect is maximal and remarkably symmetric, with a characteristic angle \( \alpha_N \approx 0.712 \) (Fig.~\ref{Nernst oscillations}a), reflecting the perfect electron-hole symmetry. In doped samples, the Nernst signal develops oscillations around zero, directly mirroring the angular resonances seen in Fig. 3(c). The field dependence (Fig.~\ref{Nernst oscillations}b) is particularly telling: while the Nernst coefficient at \( \mu = 0 \) decays smoothly with increasing field, the doped case exhibits a series of sign-alternating oscillations. This sign reversal is a direct manifestation of the counter-propagating electron and hole channels switching dominance as the cyclotron radius sweeps through successive geometric resonances. These results establish the local Nernst effect as an exceptionally sensitive probe of ballistic trajectory quantization, capable of resolving individual skipping orbits without the need for quantum coherence.

Finally, we address the effect of the driving electric field. Strictly, in crossed fields, the trajectory is a cycloid. Within linear response, we neglect the perturbation's influence on trajectories that is a standard approach. Crucially, in graphene "the drift" parameter $\Delta_\text{dr} = \omega_c^{-1} \upsilon_\text{dr} / r_c = E/(\upsilon_F B)$ is energy-independent, unlike in parabolic-band systems where it varies as $E/(\sqrt{2E_F/m}\,B)$. This constancy establishes a well defined small parameter $E/(\upsilon_F B)$ by which one can neglect the drift. Moreover, graphene's large Fermi velocity makes this approximation particularly robust.

Figure~\ref{general sketch}c shows simulated trajectories without (red) and with (blue) an in-plane electric field for $\beta=E/(v_F B)=0.1$. For a disk of several microns and $B=1$ T, this corresponds to $E\approx10^3$ V/cm that is a relatively strong field. In typical experiments with $\beta\lesssim0.01$, the deviation from pure cyclotron motion is even weaker and, given finite contact widths, practically negligible.

In conclusion, we have demonstrated that a ballistic graphene disk in a weak magnetic field reveals edge currents and geometric oscillations in local thermoelectric transport. Extending this analysis to two-dimensional topological insulators, where geometric oscillations coexist with Berry-curvature-induced anomalous Nernst signals, presents a promising direction of research in mesoscopic quantum transport ~\cite{alisultanov2025thermoelectric}. Our system also offers a platform for exploring hydrodynamic electron flow characterized by frequent electron-electron collisions, current whirlpools, and negative vicinity resistance~\cite{torre2015nonlocal,bandurin2016negative} allowing geometric oscillations to be disentangled from viscous signatures such as the Gurzhi effect~\cite{gurzhi1968hydrodynamic,molenkamp1994observation}. Moreover, the disk geometry is ideally suited to study the electron–hole Aharonov–Bohm interference in real space and its influence on local thermomagnetic transport~\cite{van1998magneto}. Finally, the contrast with strain-induced chaotic dynamics in anisotropic bilayer graphene billiards~\cite{seemann2025complex} showcases how real-space confinement and momentum-space anisotropy represent distinct knobs for controlling ballistic electron flow.

\textit{Acknowledgments}. The study is supported by the Ministry of Science and Higher Education of the Russian Federation (Goszadaniye) (No. FSMG-2026-0012).

\bibliography{apssamp}

\end{document}